\documentclass[twocolumn,floatfix,showpacs,preprintnumbers,amsmath,amssymb,article]{revtex4}

\usepackage{graphics}
\usepackage{graphicx}
\usepackage{dcolumn}
\usepackage{bm}

\usepackage{amssymb}
\usepackage{perpage}	 

\MakePerPage{footnote}	 

\begin{document}

\title{Dopant metrology in advanced FinFETs}

\author{G. P. Lansbergen$^{1}$}
\author{R. Rahman$^{2}$}
\author{G.C. Tettamanzi$^{1,3}$}
\email{g.tettamanzi@unsw.edu.au}
\author{J. Verduijn$^{1,3}$}
\author{L. C. L. Hollenberg$^{4}$}
\author{G.\,Klimeck$^{5,6}$}
\author{S.\,Rogge$^{1,3}$}

\affiliation{$^{1}$Kavli Institute of Nanoscience, Delft University of Technology, Lorentzweg 1, 2628 CJ Delft, The Netherlands}
\affiliation{$^{2}$Advanced Device Technologies, Sandia National Laboratories, Albuquerque, NM 87185, USA}
\affiliation{$^{3}$ CQC2T, University of New South Wales, Sydney, NSW 2052, Australia.}
\affiliation{$^{4}$ CQC2T, School of Physics, University of Melbourne, VIC 3010, Australia}
 \affiliation{$^{5}$Network for Computational Nanotechnology, Purdue University, West Lafayette, Indiana 47907, USA}
\affiliation{$^{6}$Jet Propulsion Laboratory, California Institute of Technology, Pasadena, California 91109, USA}



\begin{abstract}
\centering
Ultra-scaled FinFET transistors bear unique fingerprint-like device-to-device differences attributed to random single impurities. This paper describes how, through correlation of experimental data with multimillion atom tight-binding simulations using the NEMO 3-D code, it is possible to identify the impurity's chemical species and determine their concentration, local electric field and depth below the Si/SiO$_{\mathrm{2}}$ interface. The ability to model the excited states rather than just the ground state is the critical component of the analysis and allows the demonstration of a new approach to atomistic impurity metrology. 

\end{abstract}

 \maketitle

\section{RECENT PROGRESS IN DONOR SPECTROSCOPY}\label{sec1.2}

Modern transistors are getting so small that it is increasingly difficult to use traditional techniques for their study and their characterisation ~\cite{sze1}. This is particularly true for the identification of the impurity chemical species present in the channel and for the quantification of their concentration. Different groups have recently investigated the effects of ultra-scaling in silicon Field Effect Transistor (FET) geometries and interesting results have emerged:


\begin{itemize}
\item The IBM Zurich group \cite{Bjo103} has shown that screening due to interface traps in ultra-scaled silicon nanowires can cause substantial increase on the ionization energy of the dopant. This result has profound implications for the design of future FET devices. 

\item M. Pierre \textit{et al} \cite{Pie133} and R. Wacquez \textit{et al} \cite{Wac193} have shown that the presence of a single dopant in the channel of a trigate FET can dramatically alter its electrical signature, even at room temperature. 

\item M. Fueschsle \textit{et al} \cite{Fue502} have studied the band structure effects on single-crystal silicon geometries for which the source, the drain and the gates are fabricated using an atomically sharp doping procedure \cite{Sch136104}. Therefore they have investigated on the consequences of scaling a device up to the single atom limit.

\item M. Tabe \textit{et al} \cite{Tab016803} have demonstrated that, in ultra scaled silicon FET devices, even in the presence of a dopant rich environment, it is possible to observe the signature of a single dopant.
\end{itemize}

\noindent Overall, all these studies have allowed a better understanding of the effects that arise due to the ultra-scaled environment, however, they have also indicated that, for the successful design of future Complementary-Metal-Oxide-Semiconductor (CMOS) devices, a substantial amount of knowledge is still missing. As an example it is not yet clear how CMOS technology will be able to overcome critical challenges such as scaling-induced variability of device characteristics \cite{Wac193,Ase153}.


\noindent As a consequence, it is of interest to discuss more in details a method that can be used to demonstrate atomic impurity metrology \cite{LanIEDM,Lan656}. In fact, through correlation of experimental data with multimillion atom simulations in NEMO 3-D, the impurity's chemical species can be identified and their concentration, local electric field and depth below the Si/SiO$_{\mathrm{2}}$ interface can be determined \cite{LanIEDM,Lan656}. Furthermore, the extension of the dopants in the source/drain (S/D) regions can be measured by spectroscopy of confined states in the channel \cite{LanIEDM,Lan656}. Lastly, the effective current distribution in the channel can be determined by means of thermionic emission theory \cite{Sel073502}. Following these lines, the goal of this paper is to introduce a new dopant metrology technique to be used for ultra-scaled CMOS devices.

\begin{figure*}[htb]
\centering
\includegraphics[width=4.84 in,height=3.44 in]{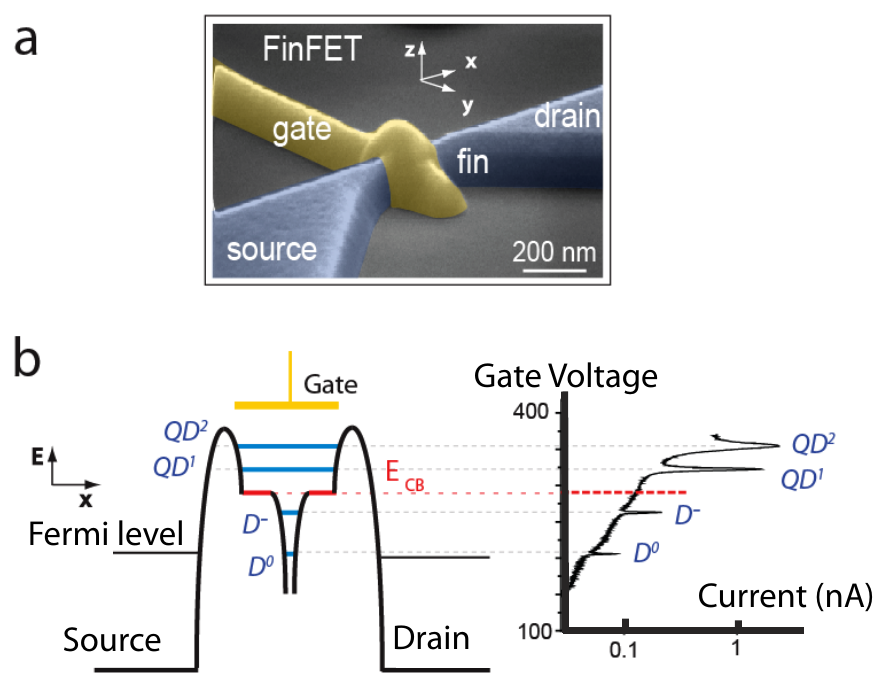}
\caption{Geometry and electrical characteristics of a single donor located 
in the channel of a FinFET device. (a) Scanning Electron Micrograph 
of a typical FinFET device. (b) Band diagram along the x-direction with the 
D$^{\mathrm{0}}$-state in resonance combined with the measured source/drain 
current versus gate voltage for a typical sample. QD$^{\mathrm{1}}$ and 
QD$^{\mathrm{2}}$ indicate resonances of a quantum dot, formed by the 
confinement provided by the corner effect and residual barriers in the 
access regions between source/drain and channel. The gate voltage where the 
band edge in the channel is aligned with the Fermi energy E$_{\mathrm{F}}$ 
in source/drain, indicated by E$_{\mathrm{CB}}$, is estimated by subtracting 
one unit of addition energy from QD$^{\mathrm{1}}$. Below the band edge, 
there are resonances ascribed to the D$^{\mathrm{0}}$ and D$^{\mathrm{-}}$ 
charge states of a single donor.}
\label{fig:Chap1Fig1}
\end{figure*}

\section{TRANSPORT BASED DOPANT METROLOGY IN ADVANCED FINFETS}\label{sec:1.3}

\noindent Direct or impurity-mediated tunnelling between source and drain competes with the thermally activated current and thereby affects the sub-threshold swing. Advanced Si MOSFET devices have shrunk to dimensions where the magnitude of the sub-threshold swing is dominated by the nature of individual impurities in the channel region. Due to the random nature of the impurity distribution, parameters like excited band levels \cite{Fue502}, ionization energy, threshold voltage and leakage current show device-to-device fluctuations \cite{Ase153,Ono102106,Kha263513,Kha223501,Tab016803,Pie133,Wac193,Bjo103}. An atomistic point of view is imperative to understand and determine the underlying donor characteristics in the channel. Several recent experiments have showed that the fingerprint of a single dopant can be identified in low-temperature transport through such devices \cite{Cal096805,Hof1261,Tab016803,Pie133,Wac193,Lan656,Sel206805}, suggesting a new method to characterise devices down to the level of a single impurity. In the context of quantum state control in quantum computing, it was shown that it is possible to model the orbital levels of a single donor in the channel of scaled FinFETs by means of multi-million atom modelling \cite{Lan656,Rah165314}. In this chapter it is shown that the aforementioned technique can be used to perform single donor mapping in advanced Si FinFETs. By carefully examining the fingerprints of isolated donors in an ensemble of devices, the chemical species, concentration, and local field of donors in the channel region can be determined in a non-intrusive fashion.

\section{DEVICES}\label{sec:1.4}

The FinFET devices used in this study consist of crystalline silicon wires (fins) with large contacts patterned by 193 nm optical-lithography and dry etching from Silicon-On-Insulator, see Fig.~\ref{fig:Chap1Fig1} (a). After a boron channel implantation, a 100 nm polycrystalline silicon layer was deposited on top of a nitrided oxide (1.4 nm equivalent SiO$_{\mathrm{2}}$ oxide thickness). A phosphorus (P) implant was used for pre-doping and the structure was subsequently patterned using an oxide hard mask to form a narrow gate. Next, high-angle arsenic (As) implantations were used for source or drain extensions, while the channel was protected by the gate and 50 nm wide nitride spacers and remained p type. Finally, As and P implants and a NiSi metallic silicide were used to complete the source or drain electrodes. The samples described in this chapter all have a gate length of 60 nm.

Transport measurements are performed on an ensemble of devices at a temperature of 4 K and a search for the fingerprints of isolated donors is performed. These single donors are located in or near the active cross-section of the channel, i.e. the cross-section of the FET body where the potential is lowest and the electrical transport thus takes place \cite{Sel073502}. Large electric fields induced by the gate or even corner effects can reduce the active cross-section to dimensions much smaller than the FET body \cite{Sel073502}. Furthermore, corner effects play a major role in these devices, as indicated by an active cross-section of only 4 $nm^2$, determined by thermionic transport measurements \cite{Sel073502}.


A single donor's fingerprint is characterised by a pair of resonances in the source-drain current, I$_{\mathrm{SD}}$, versus gate voltage, V$_{\mathrm{G}}$, characteristics at low V$_{\mathrm{SD}}$ (Fig.~\ref{fig:Chap1Fig1} (b)). The positions of a pair of resonances in V$_{\mathrm{G}}$ are an indication

\begin{figure*}[htb]
\centering
\includegraphics[width=3.25 in,height= 6.6 in]{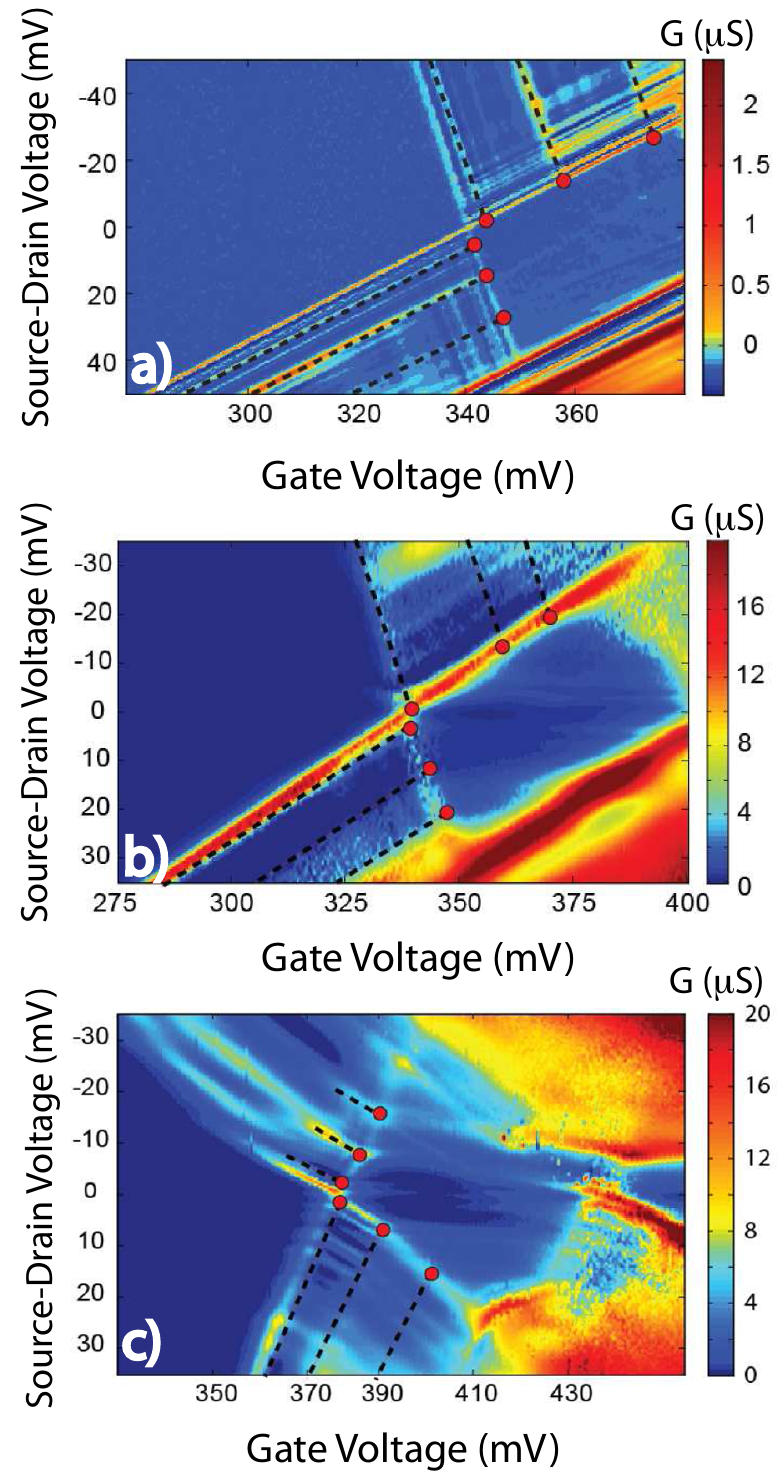}
\caption{Source/drain differential conductance of the D$^{\mathrm{0}}$ 
charge state as a function of bias voltage and gate voltage of typical 
single donor FinFET devices. The excited states, indicated by the black
dashed lines, form the fingerprint by which the donor 
properties can be identified. The dots are a direct indication for the energy of these 
states ($E_{i}$ = $eV_{\mathrm{SD}}$, with $E_{i}$ the i-th excited state and e the 
unit charge.) a) Sample 13G14: Excited states are observed at 3.5, 15.5 and 
26.4 meV b) Sample 10G16: Excited states are observed at 2, 15 an 23 meV. c) 
Sample GLJ17: Excited states are observed at 2, 7.7 and 15.5 meV.}
\label{fig:Chap1Fig2}
\end{figure*}

\noindent of the energy of the one-electron (D$^{\mathrm{0}})$ and two electron charge states (D$^{\mathrm{-}})$. A large quantum dot present in the channel (with charge states indicated by QD$^{\mathrm{1}}$ and QD$^{\mathrm{2}}$) is also observed, which allows a rough determination of the position of the band edge in the active area. While the quantum dot in the channel is almost always found, only about one out of seven devices shows the fingerprint of a donors. The identification of the resonances of the donor is based on the determined binding energy, charging energy and the odd-even spin filling \cite{Sel206805}. Next, the excited energy levels of the one-electron (D$^{\mathrm{0}})$-state are determined by sweeping both the V$_{\mathrm{G}}$ and V$_{\mathrm{SD}}$ biases and measuring the differential conductance ($dI_{SD}/dV_{SD}$) in the appropriate bias space, see Fig. ~\ref{fig:Chap1Fig2}. In this so-called stability diagram the typical diamond-shaped region associated with Coulomb-blocked transport between the D$^{\mathrm{0}}$ and D$^{\mathrm{-}}$ states is observed. The total electronic transport in the conducting regions increases as an excited level of

\begin{figure*}[htb]
\centering
\includegraphics[width=6.17 in,height=3.87 in]{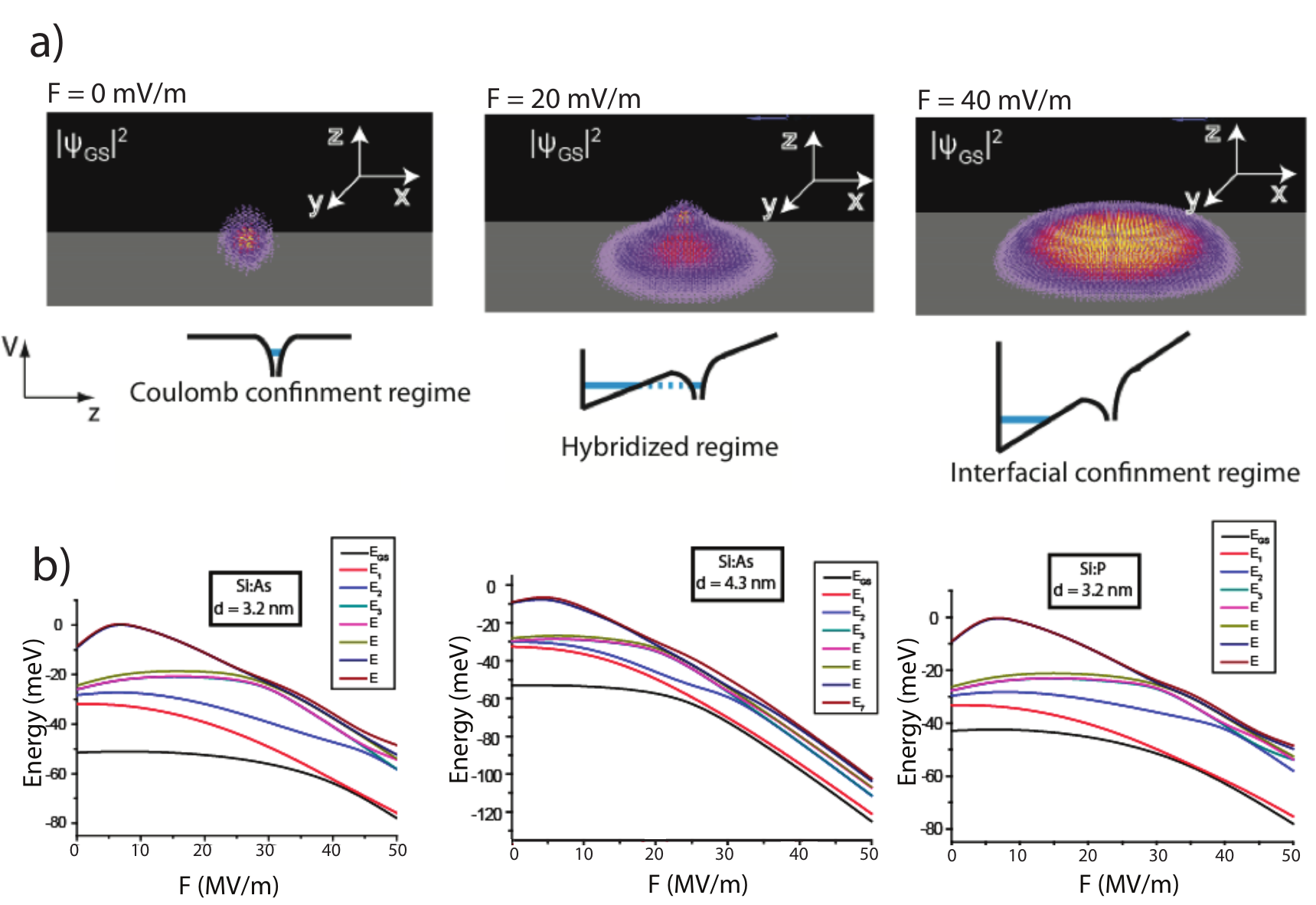}
\caption{a) Calculated wave-function density of an As donor with 
d $=$ 4.3 nm for three different fields. The gray plane indicates the 
SiO$_{\mathrm{2}}$-interface. From low-fields (where the donor has a 
bulk-like spectrum) to high fields, the donor electron makes a transition 
from being localised on the donor to being localised at the silicon interface. b) First eight eigen-levels of an As donor 3.2 and 4.3 nm below the 
interface and a P donor 3.2 nm below the SiO$_{\mathrm{2}}$ interface as a 
function of electric field (F) calculated in a tight-binding model (NEMO 
3-D). Note that we measure excited states relative to ground state (lower black 
line in this graph.)}
\label{fig:Chap1Fig3}
\end{figure*}

\noindent the D$^{\mathrm{0}}$-state enters the bias window defined by source/drain, giving the stability diagram its characteristic pattern \cite{Sel206805,Sel073502} indicated by the dashed black lines. The red dots indicate the combinations of V$_{\mathrm{SD}}$ and V$_{\mathrm{G}}$ where the ground state is at the Fermi energy of the drain and an excited state is at the Fermi energy of the source. It is the bias voltage V$_{\mathrm{SD}}$ in this combination that is a direct measure for the energy of the excited state (eV$_{\mathrm{SD,N\thinspace \thinspace }}=$ E$_{\mathrm{N}})$, where E$_{\mathrm{N}}$ is the energy relative to the ground state and N is the level index). The excited states as determined in this fashion are depicted in Table I. These levels are not bulk-like but heavily influenced by the local electric field and the nearby interface \cite{Smi193302,Cal096802,Rah165314}.

\noindent Finally, the measured level spectrum is compared to a multimillion atom tight-binding NEMO 3-D calculation \cite{Kli2090} of the system taking two possible chemical species, As and P, into account. NEMO 3-D solves for the eigenvalues of the single electron Schr\"{o}dinger all band equation in a tight-binding approach \cite{Kli601}. The NEMO simulation package is based on about 14 years of development at Texas Instruments, NASA Jet Propulsion Laboratory, and Purdue University \cite{Kli601,Kli2090}. Each atom is explicitly represented and the electronic structure of the valence electrons is represented by ten sp3d5s* orbitals. Spin can be included explicitly into the basis by doubling the number of orbitals. Spins are coupled through spin-orbit coupling resulting in accurate valence band states. The five d orbitals help shape the curvature of the conduction bands to achieve appropriate masses and symmetries at X and L. The tight-binding parameters are tuned to reproduce the bulk silicon properties under various strain conditions faithfully. For systems where the primary interest is in the conduction band properties and if no magnetic fields need to be considered, spin can safely be ignored without any significant loss of accuracy. Effects due to crystal symmetry, strain, local disorder, and interfaces can be explicitly included in the model through direct atomic representation.

\noindent The single impurity states are modelled with a simple Coulomb potential away from the impurity site and a central on-site core correction to match experimentally observed bulk-like impurity energies. The simulation domain for a bulk-like single impurity must be large enough such that the hard wall boundary conditions imposed by the finite simulation domain are not felt by the central impurity. With a simulation domain of 30.4x30.4x30.4nm$^{\mathrm{3}}$ corresponding to about 1.4 million silicon atoms the impurity eigenstates move less than 1 $\mu $eV with further domain size increases. A critical modelling capability here is the need to be able to compute

\begin{table*}[htbp]
\begin{center}
\begin{tabular}{|l|l|l|l|l|l|l|p{31pt}|}
\hline
Device \newline
& 
& 
E1 (meV)& 
E2 \newline
(meV)& 
E3 \newline
(meV)& 
d (nm)& 
F (MV/m)& 
s \newline
(meV) \\
\hline
10G16& 
Exp& 
2& 
15& 
23& 
\multicolumn{3}{|p{94pt}|}{} \\
\hline
& 
TB& 
2.2& 
15.6& 
23.0& 
3.3& 
37.3& 
0.59 \\
\hline
11G14& 
Exp& 
4.5& 
13.5& 
25& 
\multicolumn{3}{|p{94pt}|}{} \\
\hline
& 
TB& 
4.5& 
13.5& 
25.0& 
3.5& 
31.6& 
0.04 \\
\hline
13G14& 
Exp& 
3.5& 
15.5& 
26.4& 
\multicolumn{3}{|p{94pt}|}{} \\
\hline
& 
TB& 
3.6& 
15.7& 
26.3& 
3.2& 
35.4& 
0.17 \\
\hline
HSJ18& 
Exp& 
5& 
10& 
21.5& 
\multicolumn{3}{|p{94pt}|}{} \\
\hline
& 
TB& 
4.5& 
9.9& 
21.8& 
4.1& 
26.1& 
0.63 \\
\hline
GLG14& 
Exp& 
1.3& 
10& 
13.2& 
\multicolumn{3}{|p{94pt}|}{} \\
\hline
& 
TB& 
1.3& 
10& 
12.4& 
5.2& 
23.1& 
0.28 \\
\hline
GLJ17& 
Exp& 
2& 
7.7& 
15.5& 
\multicolumn{3}{|p{94pt}|}{} \\
\hline
& 
TB& 
1.3& 
7.7& 
15.8& 
4.9& 
21.9& 
0.77 \\
\hline
\end{tabular}
\caption{First three measured excited states of each sample (see for example Fig. 2) versus the best fit to the NEMO 3-D model (as depicted in Fig. 3). The fit yields a unique combination of (F, d) for each single donor device. The measurement error for each level is estimated to be around 0.5 meV.}
\label{tab1}
\end{center}
\end{table*}

\noindent reliably the ground states as well as the excited states of the impurity system, as that is a significant component of the impurity fingerprint. Fig.~\ref{fig:Chap1Fig3} (b) shows typical eigenstate spectra for an As donor (two donor depths) and a P donor as a function of electric field. Three electric field regimes can be distinguished (Fig.~\ref{fig:Chap1Fig3} (a)). At the low field limit (F $\sim$ 0 mV/m) the spectrum of a bulk As donor is obtained. In the high field limit (F $\sim$ 40 MV/m) the electron is pulled into a triangular well formed at the interface and the donor is ionized \cite{Smi193302,Cal096802,Lan656}. In the cross-over regime (F $\sim$ 20 MV/m) the electron is de-localised over the donor- and triangular well potential and the level spectrum consists of levels associated with the donors, levels associated with the triangular well at the gate interface (formed by the local field) and hybridised combinations of the two \cite{Cal096802,Lan656,Rah165314}.

The measured level spectra are least-square fitted to a calculation over a sufficiently sized region of F-d (field and donor depth parameter space). At least three excited levels per donor are taken into account to make the fit over-determined. The fitting procedure is performed for two different species of donor atoms, As and P, which were both used in the fabrication process. The concentration of donors in or near the active area of the FET channel can be estimated by comparing how many times a single donor is identified with the relevant volume where donors can be found.

\section{RESULTS}\label{sec:2.3}

Of the forty-two devices that have been examined, six have been found to exhibit the fingerprint of a single donor in the transport characteristics. These devices were subsequently measured carefully and the D$^{\mathrm{0}}$ level spectrum was fitted. The quality of the fits, indicated by $\chi $2, across the six samples is 0.92 ($\chi $2 \textless 1 means a good fit) assuming As donors. For P donors, we find a $\chi $2 of 10.22 (although two of the six are comparable in quality.) Based on the fits, the donors active in the devices are assumed to be As. The (over-determined) fit furthermore yields a unique combination of (F, d) for each single donor device, as shown in Table I, together with the measured level spectra, and their fits.

Finally, it is also possible to obtain the donor concentration. As mentioned before, the active cross-section of the devices are heavily reduced by a corner effect \cite{Sel073502}. The identified donors are either in or near these corners, and from the least-square fit we actually know the donor depths range from $\sim$ 3 to 6 nm below the Si/SiO$_{\mathrm{2}}$ interface, see Table I. The part of the FinFET channel where donors are found is a volume spanned by the donor depths and the gate length (60 nm), and a single donor device is found in average only in one out of seven devices that are measured. This means that for each $\sim$ 1000 $nm^3$ of measured material it is possible to found one As dopant, and, as a consequence, a local As concentration of about 10$^{\mathrm{18}}$cm$^{\mathrm{-3}}$ can be extrapolated.


\section{CONCLUSIONS}\label{sec:2.3}

In this paper a novel method for dopant metrology is introduced. The excellent quantitative agreement between the measured and modelled level spectra gives an indication of the high level of confidence of this method to determine the chemical species and local field of single impurities in silicon FinFET transistors. Furthermore, the local concentration of donors near the active cross-section of the FinFET can be estimated. The present method offers opportunities for non-invasive characterisation down to the level of a single donor and could be a future tool in the guidance of device processing. This is especially true for such CMOS devices for which variability problems are increasing dramatically with ultra-scaling and this justifies the importance of the method described here.

\section{ACKNOWLEDGEMENTS}

G.P.L., G.C.T., J.V. and S.R. acknowledge financial support from the EC FP7 FET-proactive NanoICT projects MOLOC (215750) and AFSiD (214989) and the Dutch Fundamenteel Onderzoek der Materie FOM. G.C.T., J.V., L.C.L.H. and S.R. acknowledge financial support from the ARC-CQC2T (project number CE110001027).
L.C.L.H. acknowledges the US National Security Agency and the US Army Research Office under contract number W911NF-08-1-0527. R.R. and G. K. acknowledge financial support from U.S. National Security Agency (NSA) and the Army Research Office (ARO) under Contract No. W911NF-04-1-0290. The work at Purdue and JPL is supported by grants from the Army Research Office. The work at the Jet Propulsion Laboratory, California Institute of Technology, is supported by grants from the National Aeronautics and Space Administration. R.R. acknowledges the financial support from Sandia, which is a multiprogram laboratory operated by SandiaCorporation, a Lockheed Martin Company, for the United States Department of EnergyÕs National Nuclear Security Administration under Contract No. DE-AC04-94AL85000.

\newpage

\bibliography{bib}

\end{document}